\documentclass[notoc]{JHEP3}
\usepackage[pdftex]{graphicx}
\usepackage{amssymb}
\preprint{COLO-HEP-544}
\title{Robustness of Sound Speed and Jet Quenching for Gauge/Gravity Models of Hot QCD}
\author{Oliver DeWolfe \\ Department of Physics 390 UCB \\
University of Colorado\\ Boulder, CO 80309, U.S.A. \\E-mail:
  \email{oliver.dewolfe@colorado.edu}}
\author{Christopher Rosen \\ Department of Physics 390 UCB \\
University of Colorado\\ Boulder, CO 80309, U.S.A.  \\E-mail:
  \email{christopher.rosen@colorado.edu}}
\abstract{We probe the effectiveness and robustness of a simple gauge/gravity dual model of the QCD fireball that breaks conformal symmetry by constructing a family of similar geometries that solve the scalar/gravity equations of motion.  This family has two parameters, one of which is associated to the temperature.  We calculate two quantities, the speed of sound and the jet-quenching parameter.  We find the speed of sound to be universal and robust over all the geometries when appropriate units are used, while the jet-quenching parameter varies significantly away from the conformal limit. We note that the overall structure of the jet-quenching depends strongly on whether the running scalar is the dilaton or not.  We also discuss the variation of the scalar potential over our family of solutions, and truncate our results to where the associated error is small.}

\begin{document}
\section{Introduction}
The dual description of hot, strongly coupled gauge theories in terms
of weakly coupled supergravity in a curved spacetime is one of the
greatest surprises of the last dozen years. This correspondence is
tantalizingly close to offering a theoretical framework by which one
might hope to understand the dense matter
produced at the Relativistic Heavy Ion Collider (RHIC) --- after all,
with collision energies around 40 TeV the produced plasma is certainly
``hot'' (on the order of a couple hundred MeV), and the quantifiable
successes of hydrodynamic models applied to RHIC flow phenomena
suggest that the matter may indeed be ``strongly coupled''.

Unfortunately, it is not yet known {\it what} curved spacetime we
should study in order to calculate in quantum chromodynamics. The
simplest examples of the gauge/gravity correspondence, 
like the one
between $\mathcal{N} = 4$ Super Yang-Mills and
$AdS_5 \times S^5$, often exhibit decidedly un-QCD like
characteristics. These include supersymmetry, exact conformal symmetry and an absence of degrees of freedom transforming under the fundamental representation of the gauge group.

In light of this somewhat bothersome obstacle, there remain several
options for applying these string theoretic
techniques to hot QCD matter. First, one could continue calculating in
$AdS_5\times S^5$ and hope that the results obtained do not depend
much on the details of the geometry. Such is the case for the
ratio of the viscosity to entropy density obtained in
\cite{son}, in which the specifics of the geometry (and hence the
corresponding gauge theory) ``divide out'',
leaving a result universal over a broad class of theories
(for possible exceptions, see
\cite{Kats:2007mq, Brigante:2007nu, adams, Buchel:2008vz}). Such results are
probably
rare, but highly desirable --- their indifference to the details of the theory
imply that if QCD has a gravitational dual, they hold there as
well. 
However, since
such universality can at best hold for some subset of all
properties of hot QCD, it is useful to think about looking further. 

Alternatively, one may attempt to incorporate more QCD-like features
into the correspondence (see for example \cite{Gubser:2009md,Erdmenger:2007cm,Mateos:2007ay} and references therein). The ideal situation, of course, is to find an
exact solution to string theory or supergravity with more realistic
features, such as non-conformality, less supersymmetry or fundamental
matter.  (The {\em most} ideal situation would be to find an exact
dual to QCD, but this seems at present too much too hope for.)  A
difficulty with this important path is that only a very small number
of exact string or supergravity solutions at finite temperature with
the appropriate features are known; for example, the ${\cal N} = 2^*$
model \cite{buchel,Buchel:2008uu,Buchel:2007vy} and the Sakai-Sugimoto model \cite{sak-sug, Sakai:2005yt, Aharony:2006da, Parnachev:2006dn}.
Much important work has been done in this direction.

Another possibility, however,  is to sidestep the scarcity and complexity of exact solutions by simply postulating a spacetime geometry with the desired properties, preferably a simple one, without explicit knowledge of what matter content (if any) renders this background a solution of the supergravity equations.  For example, the metric of Kajantie, Tahkokallio and Yee \cite{Kajantie} breaks conformal symmetry by introducing a simple dimensionful parameter in a warp factor while simultaneously turning on a scalar field.  This metric was then explored by Liu, Rajagopal and Shi \cite{LRS}, and its consequences for a number of dynamical properties in the QCD fireball were worked out.
If such models can yield useful results, a great deal of effort obtaining exact solutions can be circumvented.

Even with such a model, however, it is important
to think carefully about the utility of results obtained via gauge/gravity duals. 
Although the model may move closer to having QCD-like properties, it is still not QCD, and whatever quantities one calculates are again most useful if they do not vary much over the class of spacetimes with whatever properties were imposed.  For example, any
result obtained for an arbitrary hot plasma lacking conformal
symmetry that is {\em not} QCD is only useful for understanding QCD in
as far as the result is not sensitive to which non-conformal plasma is
considered. 
Put another way, ``useful'' results for non-conformal hot plasmas only
need to be universal across the subset of gravity duals that break
conformal symmetry.
This idea is discussed at length in \cite{LRS}. 

Thus one hopes to capture the properties of a whole class of gauge theories, ideally including QCD, by constructing a simple model with one or two desired properties.  
Two natural questions then arise.  First,
are any pathologies introduced into the gauge theory behavior by taking a gravity dual that is not known to solve any equations of motion?  Even if one fully trusts the existence of the gauge/gravity correspondence, it only promises consistent and physical dual behavior for backgrounds that are genuinely consistent solutions to a theory of quantum gravity.  It is possible that backgrounds that exist in the ``gravity swampland", without any extension to a consistent theory of gravity, would predict unphysical behavior.
And second, how robust are such models in calculating physical quantities of interest?

To address these issues,
in this paper we construct gravity solutions of varying temperature closely analogous to the solutions of \cite{Kajantie}, with the same dimensionful parameter in the four-dimensional part of the metric, but which are known to solve the equations of motion for a single scalar field with a potential coupled to gravity \cite{Gubser1, Gubser2}; similar scalar/gravity models not designed to directly make contact with \cite{Kajantie} include \cite{Gubser1, Gursoy:2008bu,Gursoy:2008za}.  These solutions are necessarily more complicated than those of \cite{Kajantie}.
Moreover, these solutions come in a family parameterized not just by the temperature, but by an additional dimensionless quantity, here called $\alpha$.
We then study two properties of our metrics: a bulk property, the speed of sound, which we compare to the speed of sound calculated directly from the model \cite{Kajantie}, and the jet-quenching parameter $\hat{q}$, which we compare to the results of \cite{LRS}.  Gratifyingly, we find broad agreement, suggesting that the use of the simpler toy model not known to satisfy equations of motion is
not problematic
in this case.

Moreover, we are able to normalize the temperature scale such that the speed of sound for our family of solutions has the same dependence on temperature regardless of the additional dimensionless parameter $\alpha$.  However, we find that in this physical scale, the form of the jet-quenching parameter varies with temperature differently depending on the value of $\alpha$.  This suggests that 
our geometries
may be quite robust in predicting a speed of sound; but on the other hand 
may not be as robust for calculating the jet-quenching parameter.

The solutions we construct, although more complicated than those of
\cite{Kajantie}, share the same warp factor for the four-dimensional
part of the metric.  It is not obvious in general that a whole class
of solutions with different temperatures should exist with such a
common warp factor.  Indeed, we find that the solutions we 
obtained for varying temperatures have slightly different values of the scalar potential.  If the potential varied wildly between solutions, there would be no sense in which we could think of them as states of different temperature in the same theory.  We show, however, that for the region of physical interest, the change in the potential is small compared to the potential itself, and thus the associated errors are small.  This places a limit on the validity of our approach, where the variation of the potential becomes significant.

Finally,
in the model of \cite{Kajantie}, it was assumed that the one scalar field present was the dilaton, which affects the string metric perceived by the worldsheet in calculations of quantities such as the jet-quenching parameter $\hat{q}$.  {\em A priori}, the single-scalar solutions we construct are agnostic as to whether the scalar field is the dilaton or not.  We compare the calculation of $\hat{q}$ between the two cases and find dramatically different behavior with temperature, in one case rising and in the other falling.
We suggest that 
the evolution of the jet quenching parameter with temperature may provide a way of distinguishing
whether the spacetime dual to QCD possesses a running dilaton, and hence 
potentially a strongly-coupled
region somewhere in its extent.

We construct our family of solutions in section~\ref{review}.
After reviewing
the 
conformal symmetry-breaking
model of \cite{Kajantie}
in section~\ref{sec:KTY},
we detail our solutions in section~\ref{sec:Solns}, and discuss the variation of the scalar potential in section~\ref{sec:Potential}.
In section \ref{sec:sound}, we
calculate the speed of sound both for the model of \cite{Kajantie} and for our solutions, and show how with an appropriate choice of temperature scale, all models have the same functional form for the sound speed.  In section~\ref{sec:jet} we calculate the jet-quenching parameter $\hat{q}$ for our models and compare to the result of \cite{LRS}, using both the temperature scale of \cite{LRS} and the units where the speed of sound is universal.  We also compare the behavior of $\hat{q}$ when the scalar field is treated as the dilaton to when it is not.  We discuss our results and conclude in section~\ref{sec:Conclusion}.

\section{Probing the Plasma: Models}\label{review}

In an effort to measure how useful a result computed in the 
gauge/gravity correspondence is to RHIC physics, the
authors of \cite{LRS} offer a pragmatic routine. First, choose a
simple metric that preserves all the desired spacetime symmetries and
reduces to $AdS_5$ in the UV, but breaks conformal
invariance. Effectively, this allows one to ``turn on'' a feature of
QCD not typically present in pure AdS/CFT. Next, one uses this metric as
input to computation in the gauge/gravity correspondence. By studying
how the results depend on the conformal symmetry breaking parameter,
one hopes to measure the ``robustness'' of some result to the shift
from model to QCD. 

In this section, we review the model of Kajantie, Tahkokallio and Yee (KTY) \cite{Kajantie} employed by Liu, Rajagopal and Shi \cite{LRS} to undertake this program, and then construct a class of backgrounds sharing the same conformal-symmetry breaking parameter that explicitly satisfy the equations of motion for one scalar coupled to gravity.

\subsection{KTY model}
\label{sec:KTY}

Let us review the model of \cite{Kajantie}. We use conventions where the action of the gravity/scalar system reads
\begin{equation}\label{eq:action}
\mathcal{S} = \frac{1}{16\pi G_5}\int d^5x\sqrt{-g}\left(R -
  \frac{1}{2}(\partial_\mu\Phi)^2 -V(\Phi) \right) \,.
\end{equation}
In these conventions, the KTY model takes the form
\begin{eqnarray}
\label{KTYmetric}
ds^2 = {R^2 \over z^2} e^{2f(z)} \left[ - h(z) dt^2 + d\vec{x}^2 + {dz^2 \over h(z)} \right]\,,
\end{eqnarray}
with 
\begin{eqnarray}
\label{KTYwarp}
2f(z) = - c z^2 \,, \quad \quad h(z) = 1 - {z^4\over z_0^4} \,,
\end{eqnarray}
where the associated temperature is $T = 1/\pi z_0$.  The model also contains 
a running scalar field $\Phi$,
\begin{eqnarray}
\label{Scalar}
\Phi =  \sqrt{3 \over 2} \phi z^2 \,,
\end{eqnarray}
which was interpreted in \cite{Kajantie} as the 5D dilaton; we will remain neutral on whether we treat this field as the dilaton or not until section~\ref{sec:jet}.\footnote{The canonical normalization for the 5D dilaton, used in \cite{Kajantie}, is  $\Phi_5 \equiv \sqrt{3/8} \Phi$.}
If it is the dilaton, the string metric still has the form (\ref{KTYmetric}) but with
\begin{eqnarray}
2f_{str}(z) = -c z^2 + {4 \over 3} \Phi_5 = - c z^2 + \sqrt{2 \over 3} \Phi \,.
\end{eqnarray}
The metric and scalar do not solve the equations of motion of the action (\ref{eq:action}), but it was 
supposed
that some additional, otherwise-ignorable matter could be added to generate a true solution.

There are three constant parameters associated to the solution: $c$, a parameter of nonconformality appearing in the warp factor; $\phi$, the overall normalization of the running scalar; and $z_0$, which controls the temperature.   Only two are independent, however: we note that we can change coordinates:
\begin{eqnarray}
t \to \lambda t \,, \quad \quad \vec{x} \to \lambda \vec{x} \,, \quad \quad z \to \lambda z \,,
\end{eqnarray}
which is the isometry of pure AdS space (recovered in the limit $c=\phi \to 0, z_0 \to \infty$) associated to scale transformations.  The KTY metric is then unchanged if we also scale
\begin{eqnarray}
z_0 \to \lambda z_0 \,, \quad \quad c \to \lambda^{-2} c \,, \quad \quad \phi \to \lambda^{-2} \phi \,,
\end{eqnarray}
corresponding to the freedom to rescale all mass parameters simultaneously.  The two invariant dimensionless combinations can be taken to be,
\begin{eqnarray}
c z_0^2 \propto {c \over T^2} \,, \quad \quad \alpha \equiv {c \over \phi} \,.
\end{eqnarray}
In addition to the ratio of the temperature to the conformal symmetry-breaking parameter $c$, we also have the parameter $\alpha$, encoding the ratio of conformal breaking in the metric to the breaking in the scalar $\Phi$.  Thus in addition to the temperature, there is a one-parameter family of conformal-symmetry-breaking geometries.  The field theory interpretation of $\alpha$ is in general opaque.

KTY fix $\alpha$ by appealing to the behavior in the meson spectrum of a different metric intended to correspond to the low-energy effective theory on the other side of the confinement phase transition \cite{Kajantie}:
\begin{eqnarray}
\label{alphaKTY}
\alpha_{KTY} = {20 \over 49} \,.
\end{eqnarray}
We shall consider this value, but we will generally leave $\alpha$ arbitrary, and examine the resulting physics for several different choices in the vicinity of (\ref{alphaKTY}).  By considering this parameter, we will be able to quantify how much physical quantities vary as we go from one metric to another at fixed temperature, and hence get an idea of
the robustness of
our family of solutions.

\subsection{Model solving equations of motion}
\label{sec:Solns}

We consider a metric ansatz with the general form,
\begin{equation}\label{eq:genMet}
ds^2 = e^{2 A(r)}\left(-h(r)\,dt^2 + d\vec{x}^2 \right) + \frac{e^{2B(r)}}{h(r)}dz^2 \,,
\end{equation}
where $r$ is a suitable radial variable, along with a varying scalar $\Phi(r)$.  In papers by Gubser et al.,~\cite{Gubser1, Gubser2} the equations of motion for this system following from the action (\ref{eq:action}) were worked out and found to be,
\begin{eqnarray}
\label{EqnB}
A'' - A' B' + {1 \over 6} &=& 0 \,, \\
\label{Eqnh}
h'' + (4 A' - B') h' &=& 0 \,, \\
2 e^{2B} V + 6 A' h' + h (24 {A'}^2 - 1) &=& 0 \,,
\label{EqnV}
\end{eqnarray}
in a gauge where the scalar field itself is used as the radial coordinate, $r = \Phi$; we have neglected to write the scalar equation of motion which is not algebraically independent.  

In an attempt to generate a family of solutions to these equations with the same type of conformal symmetry breaking parameter as (\ref{KTYmetric}), we assume a warp factor for the four-dimensional directions of the same form as (\ref{KTYwarp}), which using (\ref{Scalar}) and the definition $\alpha \equiv c /\phi$ becomes
\begin{equation}\label{eq:A}
A(\Phi) =
\frac{1}{2}\log\left(\sqrt{\frac{3}{2}}c\frac{R^2}{\alpha}\right)-\frac{1}{2} \log\Phi-\frac{\alpha}{\sqrt{6}}\Phi \,,
\end{equation}
using $\Phi$ itself as the radial coordinate.
It is then straightforward \cite{Gubser1} to solve (\ref{EqnB}) to find
\begin{equation}\label{eq:B}
B(\Phi) = \log\left(\frac{R}{2}\right)+\frac{1+2\alpha^2}{2\alpha^2}\log\left(1+\alpha\sqrt{\frac{2}{3}}\Phi\right)-\log\Phi-\frac{1}{\alpha\sqrt{6}}\Phi \,.
\end{equation}
Note that integration constants in both $A$ and $B$ are undetermined by (\ref{EqnB}); we have fixed $A$ to agree with the KTY metric, and $B$ to reduce to $AdS$ space with matching length scale $R$ in the $\Phi \to 0$ limit.  Beyond this parameter, this family of metrics depends on the dimensionless ratio $\alpha$.

 \begin{figure}
\begin{center}
\includegraphics[width=0.6\textwidth]
{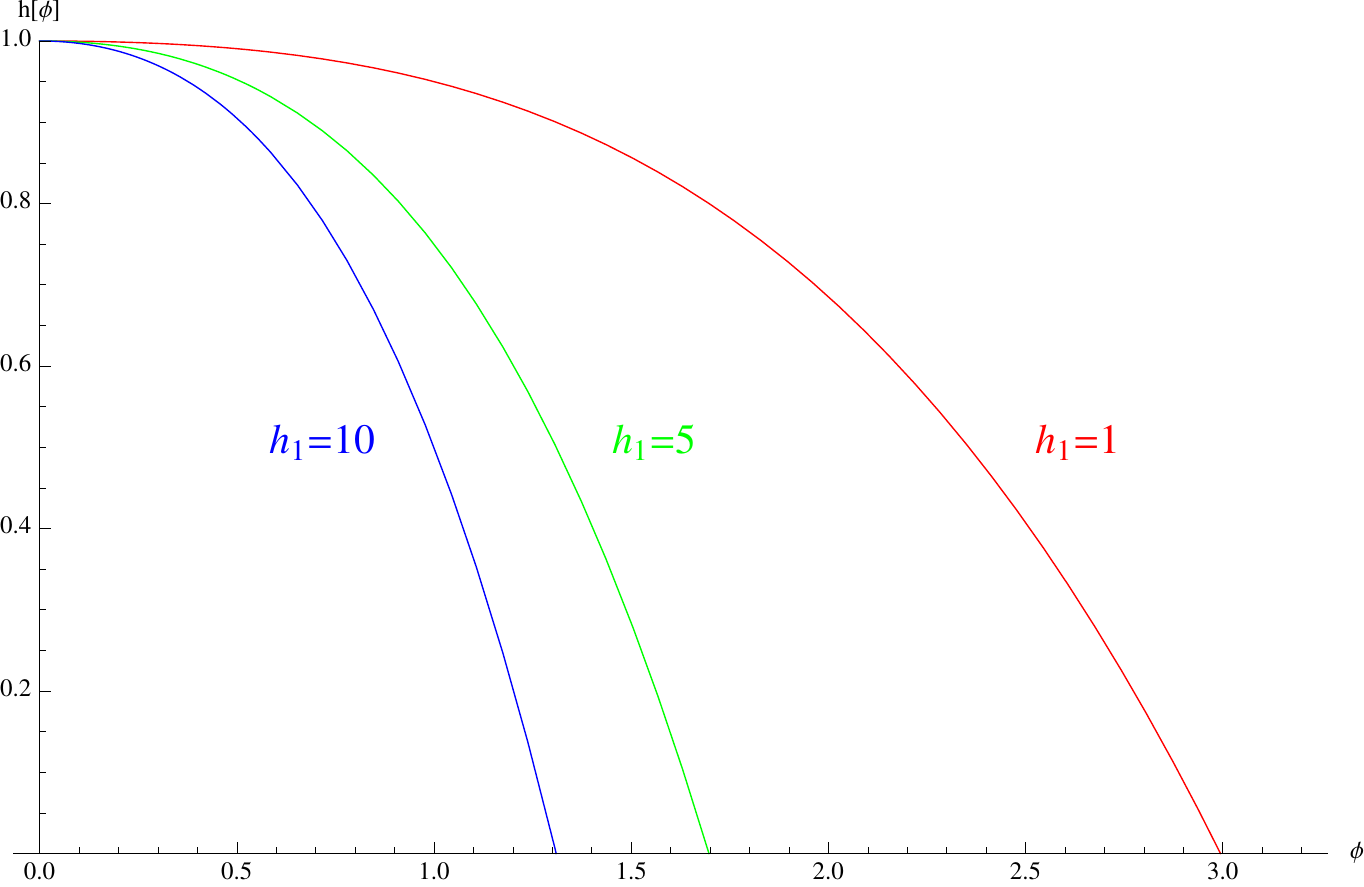}
\caption{Plots of the horizon function $h(\Phi)$ for three different values of the temperature parameter $h_1$, for $\alpha = 20/49$.
\label{fig:h}}
\end{center}
\end{figure}

We next see from equation (\ref{Eqnh}) that the horizon function $h(\Phi)$ consistent
with the equations of motion can be obtained by integrating \cite{Gubser1}
\begin{equation}\label{eq:h}
h(\Phi) = h_0 - c^2 R^3 h_1\,\int_0^\Phi d\Phi'e^{-4 A(\Phi')+B(\Phi')} \,,
\end{equation}
with the values of $A(\Phi)$ and $B(\Phi)$ obtained via (\ref{eq:A}) and (\ref{eq:B}).  Here we chose the lower bound on the integral for convenience, since any change in it can be subsumed into the constants $h_0$ and $h_1$, which are left undetermined by (\ref{Eqnh}).  We use the freedom in $h_0$ to impose $h(\Phi =0) = 1$ to reproduce the $AdS$ limit in the ultraviolet, so $h_0 = 1$.  This leaves $h_1$ as a free parameter in the set of solutions, which we will associate with the temperature.

For a black brane-type geometry 
we expect $h$ to proceed monotonically from $h=1$ at the boundary to $h=0$, which defines the horizon; we are not interested in the space beyond the horizon.  To verify that our proposed ansatz leads to actual black brane solutions, we must solve equation~(\ref{eq:h}) for $h(\Phi)$.  As it turns out, for arbitrary $\alpha$, the integral in (\ref{eq:h}) is quite complicated, but it can be evaluated  for particular choices of $\alpha$.  The result for the KTY value of $\alpha = 20/49$ can be obtained in terms of incomplete gamma functions,
\begin{eqnarray}
h(\Phi) = 1&-& A h_1  \Bigg[800\,  \Gamma \left( {4801 \over 800}, {801 \over 800} \right) - 800\, \Gamma \left( {4801 \over 800}, {267(147 + 20 \sqrt{6} \Phi) \over 39200} \right) \\ &&  -
801 \, \Gamma \left( {4001 \over 800}, {801 \over 800} \right) + 801\, \Gamma \left( {4001 \over 800}, {267(147 + 20 \sqrt{6} \Phi) \over 39200} \right) \Bigg] \,,
\nonumber
\end{eqnarray}
where $A$ is the enjoyable constant
\begin{eqnarray}
A \equiv { 163840000000000 \left(5 \over 3\right)^{1/400} 2^{1/160} e^{801/800} \over 
264116234249604801\times 89^{1/800} }  \approx 0.0017 \,.
\end{eqnarray}
A simpler choice is the nearby $\alpha = 1/2$, for which we find the polynomial result
\begin{eqnarray}
\label{h12}
h(\Phi) = 1 - {h_1 \over 4320} \left( 180 \Phi^2 + 60 \sqrt{6} \Phi^3 + 45 \Phi^4 + 2 \sqrt{6} \Phi^5 \right) \,.
\end{eqnarray}
We studied a number of choices of $\alpha$ in the vicinity $0 < \alpha < 1$,
where the integrals could be solved;
in the remaining
sections we will also plot results for $\alpha = 1/5$ and $\alpha = 4/5$.

For the particular value $\alpha = 20/49$ we plot the horizon function for three different values of $h_1$ in figure~\ref{fig:h}; we see that the function has the correct form, monotonically decreasing to the horizon, and the location of the horizon shrinks as $h_1$ grows. Similar results hold for other values of $\alpha$.  Given that our solutions do indeed possess a horizon, we can use them to study a non-conformal plasma at finite temperature.

 \begin{figure}
\begin{center}
\includegraphics[width=0.6\textwidth]
{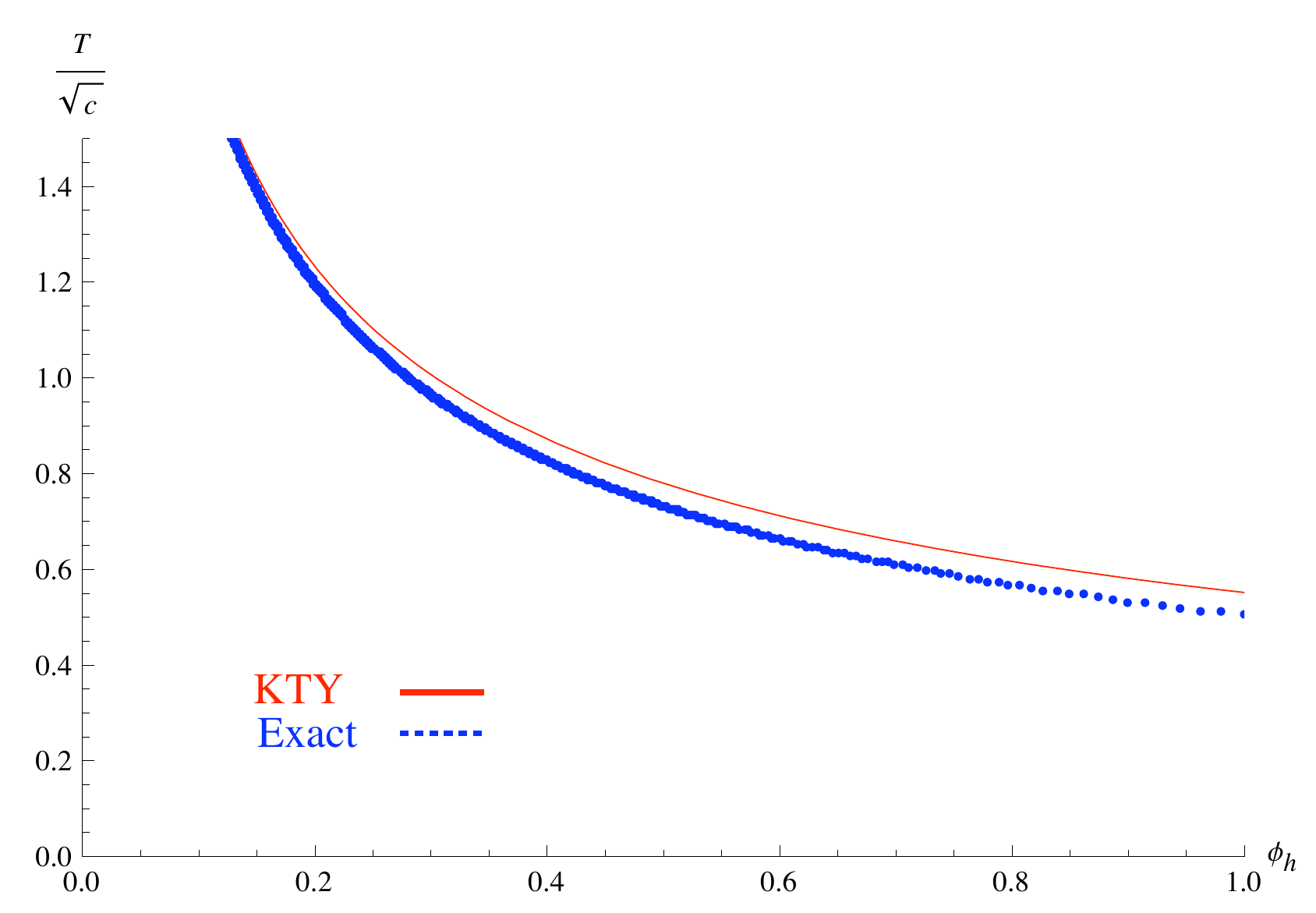}
\caption{Temperature dependence on the horizon location. Deviations
  increase as $T/\sqrt{c}$ decreases.
\label{fig:temp}}
\end{center}
\end{figure}

We would like to translate the horizon location $\Phi_h$ into a value of the temperature.  The temperature $T$ of the geometries is related to the various metric functions by 
\cite{Gubser1}
\begin{equation}\label{eq:temp}
T = \frac{e^{A(\Phi_h)-B(\Phi_h)}|h'(\Phi_h)|}{4\pi} \,,
\end{equation}
where the derivative acting on $h$ is with respect to the horizon value $\Phi_h$; note this is not the same as taking a derivative with respect to $\Phi$ and then setting $\Phi = \Phi_h$, since in general $h(\Phi)$ is also a function of the parameter $\Phi_h$.
 
 In figure \ref{fig:temp} we show the relationship between
the location of the horizon and the temperature for our solutions, and for
the KTY model, for $\alpha = 20/49$; very similar results hold for other $\alpha$ in the range $1/5 < \alpha < 1$.
To evaluate this for the model of \cite{LRS} is a
simple exercise in differentiation. For the more complicated
solution we study here, it requires a numerical
routine, where the algorithm obtains the location of the horizon for
many values of $h_1$, finds the corresponding temperature using
(\ref{eq:temp}), and compares.  The two agree in the large-$T$
(small-$c$) region
 and deviate only slightly over the range of $1/2 < T/\sqrt{c} < 1$. 
This range was argued to be of 
greatest
physical interest in \cite{LRS}, through
comparison of the thermodynamics of the KTY model to lattice
data. This small deviation is perhaps a first suggestion that
switching to explicit solutions of the equations of motion does not
create a substantial change.  For smaller temperatures they diverge
more substantially, for reasons we now describe. 

\subsection{Variation of the scalar potential}\label{sec:Potential}

\begin{figure}
\begin{center}
\includegraphics[width=0.6\textwidth]
{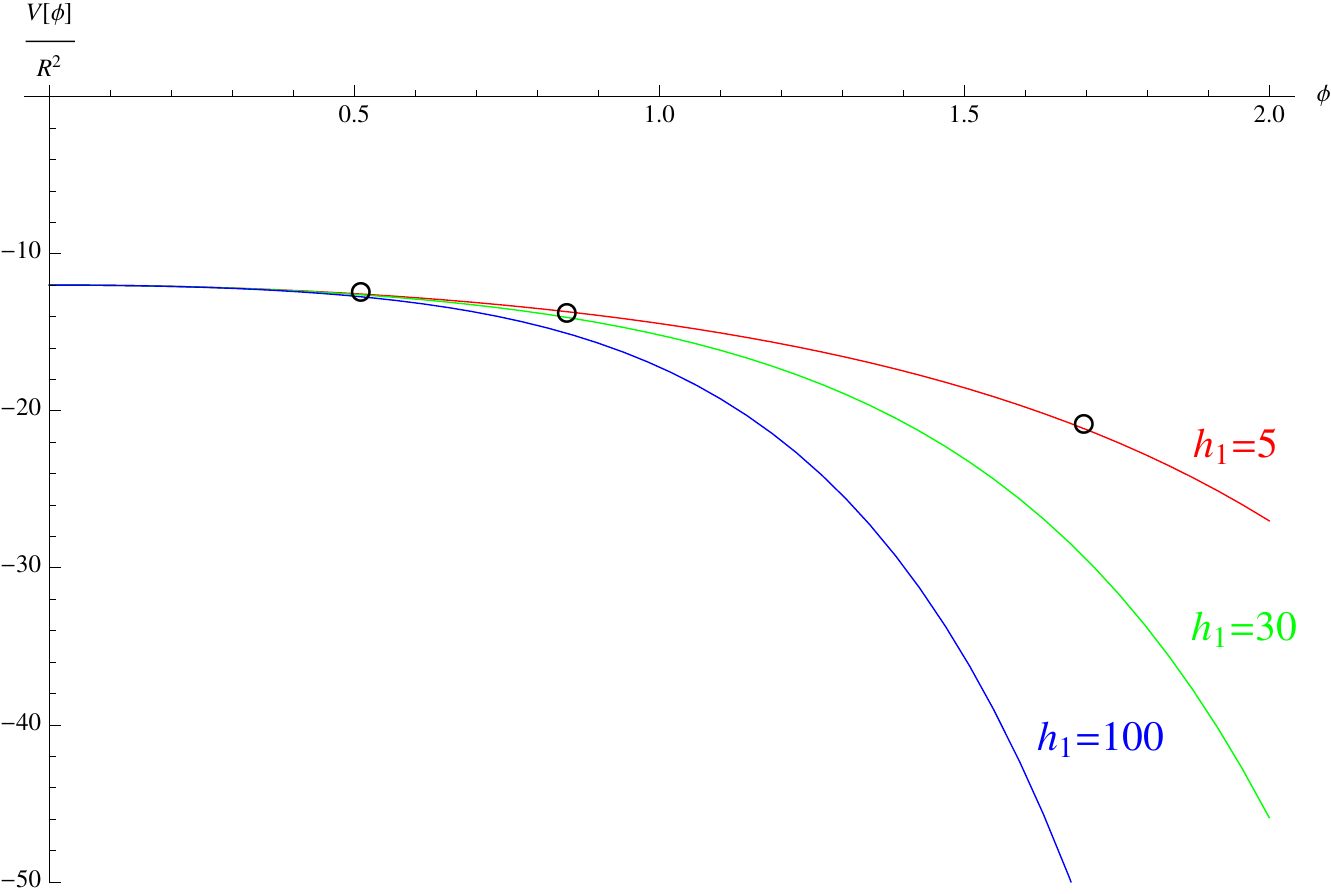}
\caption{A plot of $V(\Phi)/R^2$ for three values of the temperature, and the corresponding horizons indicated as circles.  The potential changes substantially, but most of the change is behind the horizon.
\label{fig:V}}
\end{center}
\end{figure}

So far we have discussed equations (\ref{EqnB}) and (\ref{Eqnh}), used to determine $B(\Phi)$ and $h(\Phi)$ given $A(\Phi)$.  However, we have not considered equation (\ref{EqnV}), which fixes $V(\Phi)$ in terms of the other functions.  As was noted in \cite{Gubser1}, in general this leads to a problem for solving the equations the way we have: there is no guarantee that any given set of functions $A$, $B$, and $h$ varying with temperature will lead to the same scalar potential $V(\Phi)$.  If the potential changes, then the solutions cannot be thought of as belonging to the same gauge theory dual; instead, as one varies the temperature, one is also changing parameters of the Lagrangian, and producing a vacuum of a different temperature in a {\em different} theory.

The solutions to the equations of motion we find, assuming the KTY ansatz for $A(\Phi)$ and using (\ref{EqnV}), indeed do not lead to the same potential for different temperatures.  
While it is possible that there exists {\it some} arrangement of matter
fields for which the potential remains fixed, it is in principle
difficult to find. We could also consider a more complicated ansatz, where we do not assume a simple form for $A(\Phi)$, but in addition to the issue of complexity, this defeats the purpose of comparing to the KTY model.\footnote{Solutions to the gravity/scalar system with fixed scalar potential have been studied recently in \cite{Gursoy:2008bu, Gursoy:2008za}.  These solutions indeed have different warp factors for different values of the temperature.}

Here we take a practical approach, and see how much it
matters. Examining how much the potential is varying with the temperature (that is, the location of the horizon), we find the following.   As $T$ changes, $V(0) = -12 R^2$ remains fixed, while the largest variation occurs at larger $\Phi$. However, points at sufficiently large $\Phi$ are always behind the horizon at $\Phi_h$, and hence the region of largest variation does not contribute to the physics.  As the temperature increases, the region where $V(\Phi)$ is varying substantially retreats towards $\Phi =0$, but at the same time $\Phi_h$ retreats in front of it, tending to keep the region of strong variation harmlessly behind the horizon.

We can quantify the change in $V(\Phi)$ at some point $\Phi$ as we vary the solutions over the widest 
physically relevant
range: from $T=0$ to the temperature where that point disappears behind the horizon.  We find that the variation in $V$ at a given value of the scalar increases monotonically with $\Phi$; thus for the solution at a given temperature, the point that varies the most over all temperatures to which it contributes is always the point with the largest value of $\Phi$, namely $\Phi_h$ itself.  We can thus parameterize the variation of $V$ for a solution at a particular temperature $T$, by how much the potential at the horizon point $V(\Phi_h)$ varies between that solution and the zero-temperature solution.

We have performed this analysis, and for $\alpha = 20/49$ for example we find the values
\begin{eqnarray}
{\Delta V (\Phi_h) \over V(\Phi_h)}\Bigg|_{ T / \sqrt{c} = 1} = 0.005
\,, \quad \quad
{\Delta V (\Phi_h) \over V(\Phi_h)}\Bigg|_{ T / \sqrt{c} = 1/2}= 0.04 \,,
\end{eqnarray}
over the range in $T/\sqrt{c}$ argued in \cite{LRS} to be most physically relevant.
Thus although the solutions to the equations of motion we find are not solutions for precisely the same potential, they are sufficiently similar to lead to only a small error of a couple percent  over the region we are interested in.  This error, moreover, is over the entire  possible range of temperatures; for a practical calculation comparing two solutions with similar temperatures the error will
be smaller.

  In general we find that smaller values of $T/\sqrt{c}$ have worse behavior; thus the larger the conformal symmetry-breaking, the more the effect matters.  Similar results hold for other values of $\alpha$, and as $\alpha$ increases, the ``safe" region moves further in towards smaller values of  $T/\sqrt{c}$ (and vice versa).  In the sections that follow, we arbitrarily choose a largest error in $\Delta V/V$ that we will tolerate
 of five percent,
 and truncate the results for various values of $\alpha$ each at this cut-off.  The remaining points should be robust to the variation of the potential to within a few percent.

This is a trade-off for attempting to find exact solutions that, while more complicated than the KTY model, are still relatively simple; in principle there should exist solutions with precisely the same value of $V(\Phi)$ for all temperatures, but these are in general hard to obtain, and additionally, most of these solutions will bear no particular resemblance to the KTY metric for any value of $c$.

It is also possible to vary our one other parameter, $\alpha$, as we vary $T/\sqrt{c}$, to try and mitigate the variation of the potential; that is, we can try to find the path through the two-dimensional parameter space of $T/\sqrt{c}$ and $\alpha$ that minimizes the variation of $V(\Phi)$.  It is possible to decrease the variation of the potential by several orders of magnitude by making a compensating change in $\alpha$, though one cannot cancel it precisely.  We will not attempt to systematically perform this compensation, since the error from simply changing $T/\sqrt{c}$ is already small for the region we are most interested in.

\section{The speed of sound}\label{sec:sound}

Having our solutions in hand, we would like to compare their predictions to those of the KTY model, to try and see whether the use of metrics not solving known equations of motion still leads to reasonable and robust results.  We will compare two quantities: first, a ``bulk" quantity not involving the string worldsheet, the speed of sound.  In the next section, we shall consider a worldsheet-related number, the jet-quenching parameter.

In a non-conformal plasma, one expects richer thermodynamic behavior than in the conformal case. For
example, conformal invariance fixes the speed of sound,  defined in terms of the pressure $P$ and energy density $\epsilon$ as
\begin{eqnarray}
c_s^2 \equiv {\partial P \over \partial \epsilon} \,,
\end{eqnarray}
at $c_s^2 = 1/3$. This condition comes from the tracelessness of the energy-momentum
tensor, as mandated by invariance under dilations. 
In a non-conformal plasma, on the other hand, $c_s^2$ can vary.

A convenient formula for the speed of sound is
\begin{equation}\label{eq:csGen}
c_s^2 = \frac{d\log T}{d\log s}  = \frac{d\log
  T}{d\Phi_h} \left( \frac{d\log s}{d\Phi_h}\right)^{-1} \,,
\end{equation}
valid in the limit of zero net baryon density, a reasonable
approximation in the 
heavy-ion fireball.  
Besides the temperature it requires knowledge of the entropy density $s$, which for our systems is simply proportional to the horizon area of the black brane, taking the form
\begin{equation}\label{eq:entropy}
s = \frac{e^{3A(\Phi_h)}}{4\pi G_5} \,,
\end{equation}
and since both $T$ and $s$ vary with $\Phi_h$, it is easiest for us to compute (\ref{eq:csGen}) in terms of the variation of each quantity with $\Phi_h$, as indicated.

\begin{figure}
\begin{center}
\includegraphics[width=0.6\textwidth]
{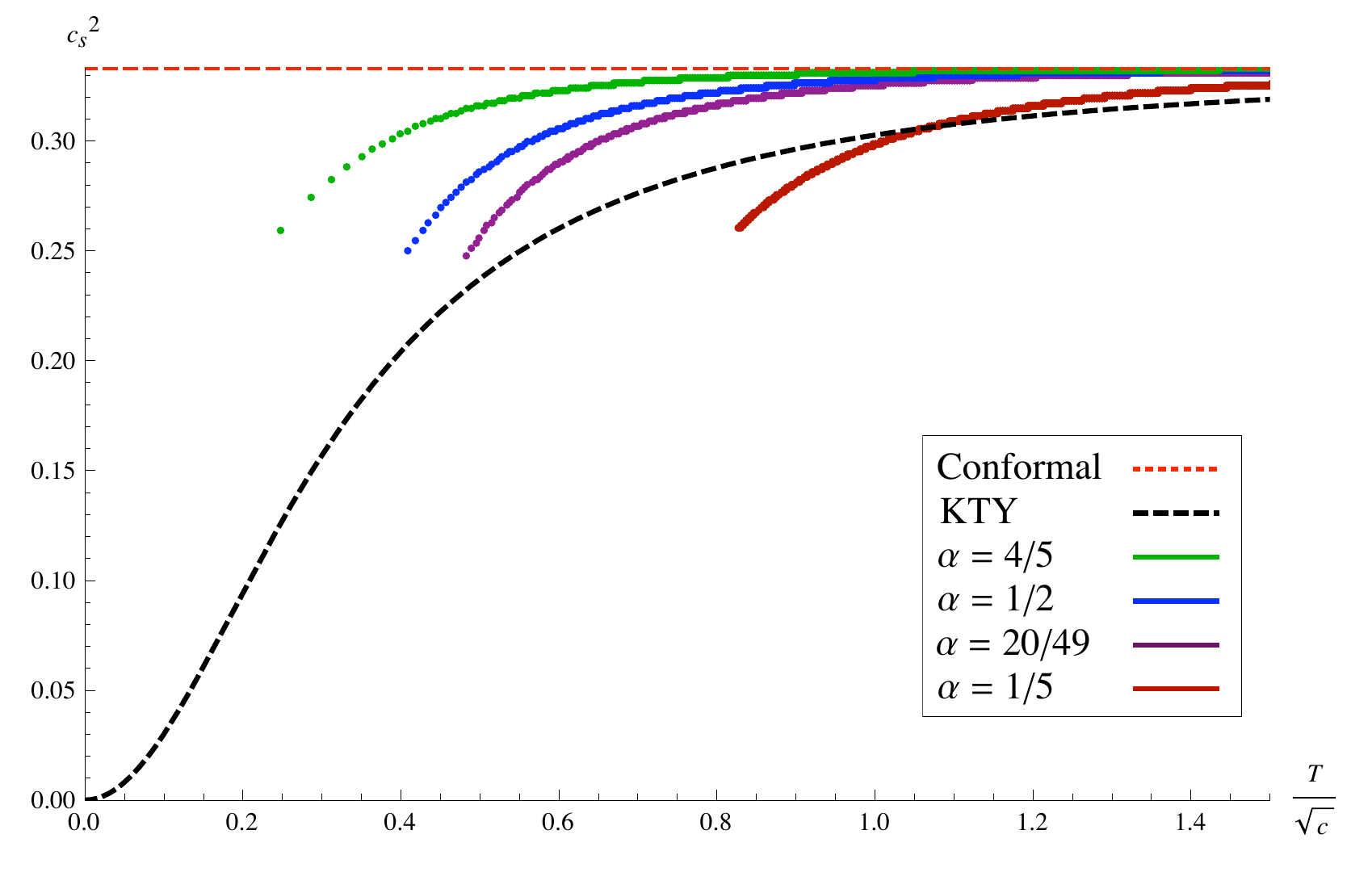}
\caption{
Behavior of the squared speed of sound with respect to the dimensionless
  quantity $T/\sqrt{c}$ plotted for several values of the
  parameter $\alpha$. Also shown is the conformal value
  ($c_s^2=1/3$), and the result given by the KTY model.
\label{fig:sound}}
\end{center}
\end{figure}

For the KTY model, the speed of sound is straightforward to evaluate:
\begin{eqnarray}
\label{KTYSound}
c_s^2 = {1 \over 3 (1 + {c \over \pi^2 T^2})} \,.
\end{eqnarray}
Note that this formula is independent of $\alpha$, which for the KTY model entered only into the scalar field and not into the metric (which controls $s$ and $T$).
The result is plotted in figure~\ref{fig:sound}.\footnote{A very similar functional form for the sound speed   was found in \cite{Andreev:2007zv}, based on the model in \cite{Andreev:2006eh} which resembles \cite{Kajantie} without a running scalar.}
This expression clearly has the right limit of $c_s^2 \to 1/3$ as we restore conformal symmetry $c \to 0$, and it monotonically decreases as the temperature gets small to approach zero in the limit of zero $T$.  This decrease 
below the conformal value
is reasonable
on physical grounds.\footnote{For example, in the simpler case of the perfect fluid $P = w \epsilon$, we have $c_s^2 = w$ and the interpolation between $c_s^2 = 1/3$ and $c_s^2 = 0$ is just the progression from ultrarelativistic to non-relativistic species.  This plasma would have no quasiparticle description, but the behavior of the speed of sound is analogous.}

For the family of solutions obtained in this paper, calculating the speed of sound in general requires a numerical
routine. The algorithm is simply to scan over many values of the horizon location $\Phi_h$ by varying $h_1$, and calculating the resulting values of the temperature and entropy.  We then compare neighboring values of $\Phi_h$ to calculate the speed of sound.  Note that because this calculation involves comparing two solutions with distinct temperature, it is particularly sensitive to the variation of the potential between those solutions; we should not expect this to be a reliable calculation in the limit that the potential variation becomes substantial.

Crucially, the
results of this calculation vary depending on the value of $\alpha$ chosen.  In 
figure \ref{fig:sound} we plot the results for a few representative
values of this parameter, along with the single result
(\ref{KTYSound}) for the KTY model.  For our results, as each model
approaches smaller values of $T/\sqrt{c}$ the error due to the
variation of the potential becomes larger.  We have therefore imposed
a somewhat arbitrary cutoff, including only 
results with $\Delta V
/V < 0.05$. 
  Following the results beyond this cutoff eventually results in pathological behavior, in particular the speed of sound squared going below zero at nonzero temperature.

For all values of $\alpha$, the speed of sound asymptotes to the
conformal value at large temperature and decreases at smaller
temperature, analogous to the KTY model.  The curvatures of our solutions are slightly different from KTY, being initially flatter and then dropping more abruptly as the temperature is lowered.  Smaller values of $\alpha$ give a lower speed of sound at
fixed temperature, while larger values  give a greater speed.
Unfortunately, the variation of the potential and the associated
curve truncation do not allow us to determine the small $T/\sqrt{c}$
behavior of the speed of sound. As discussed in \cite{Muller:2006ee},
lattice simulations for hot QCD matter predict a ``strong'' drop in
the speed of sound near the critical temperature ($T_c\sim 170$ MeV),
followed by a rise below it. Obviously, such behavior is not
present in the KTY model, which monotonically approaches $c_s^2=0$ as
$T/\sqrt{c}\to 0$. This is because we work exclusively with the
``deconfined phase'' geometry, which corresponds to $T>T_c$. For an attempt to realize this behavior in a gravity dual, see  \cite{Gubser1, Gubser:2008yx}.

We have plotted our results in terms of $T/\sqrt{c}$ in order to make comparisons with \cite{LRS}; they naturally used $c$ as a reference parameter since only it appears in the metric, which controls the quantities of interest. For full solutions to the equations of motion, however, the parameter $\phi$ feeds back into the metric as well via the parameter $\alpha$.  It is not obvious that $T/\sqrt{c}$, as opposed to $T/\sqrt{\phi}$ or any combination of the two massive parameters, is the most useful scale to employ.  In other words, our additional dimensionless parameter makes it unclear which 
units are
the natural 
ones to use.

We can use our results for the speed of sound to try and resolve this issue physically.  The speed of sound is a calculable quantity with physical consequences.  We may choose, if we can, to set our scale for each model precisely so that in these units, the same temperature always gives the same value of the speed of sound.  That is, we can seek to measure $T$ relative to whatever combination of $c$ and $\phi$ we need to make the speed of sound curve look identical for all values of $\alpha$.

\begin{figure}
\begin{center}
\includegraphics[width=0.6\textwidth]
{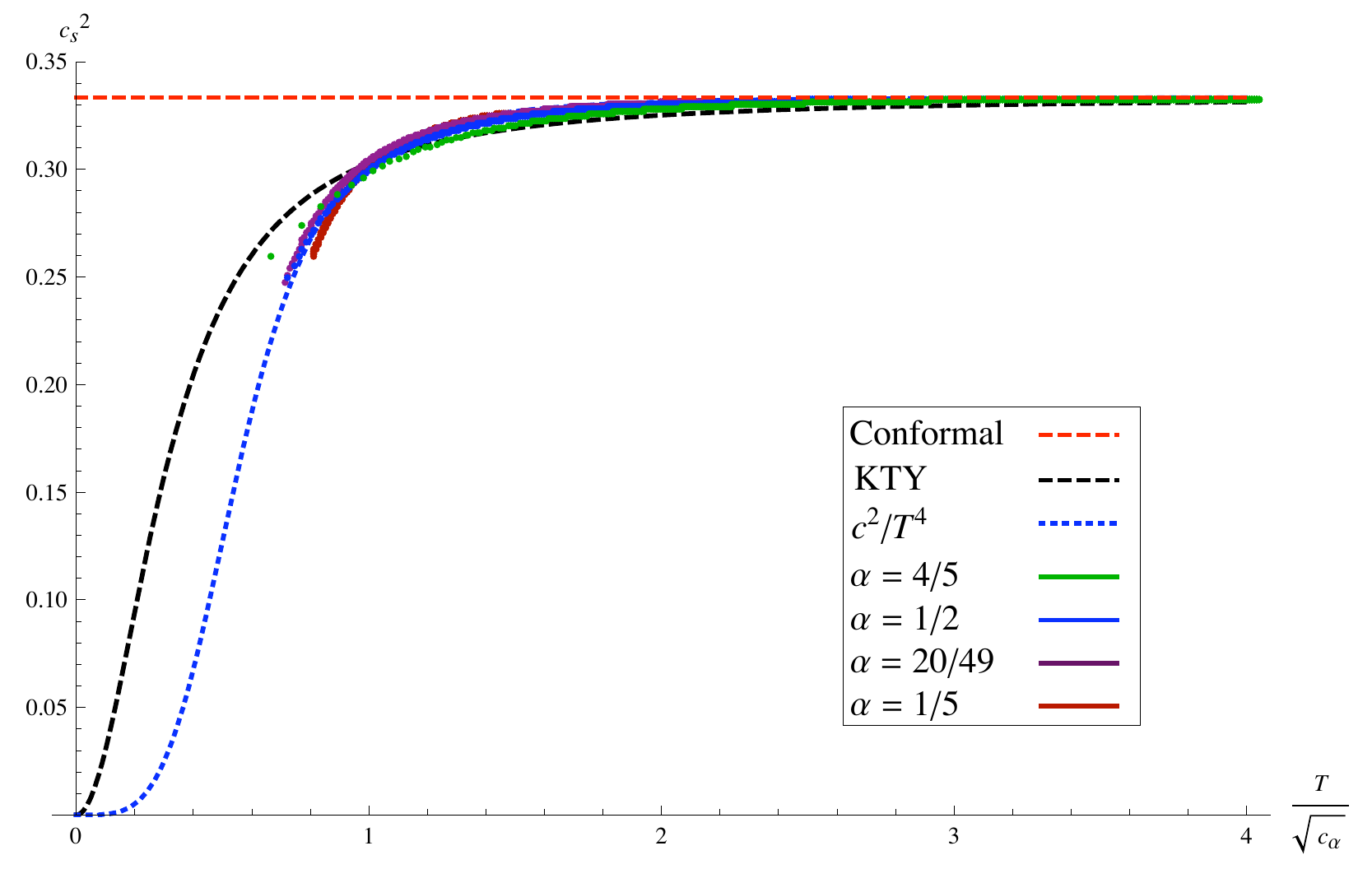}
\caption{Speed of sound curves measured in rescaled units $T/\sqrt{c_\alpha}$.
In these units, our model makes one prediction
  for the dependence of the speed of sound on the ratio of temperature
    to a dimensionful combination of the parameters $\phi$ and $c$.  We also compare to the KTY result and an analogous function depending on $T^4$. \label{fig:sound2}}
\end{center}
\end{figure}

We find that plotting the speed of sound as a
function of 
$T/\sqrt{c_\alpha}$, where $c_\alpha$ is an $\alpha$-dependent rescaling of $c$ given by $(c_{4/5}, c_{1/2}, c_{20/49}, c_{1/5}) = (0.14 c , 0.32 c, 0.46 c, 1.05 c)$ with similar scalings for other values of $\alpha$, creates a convergence between the various curves of our model, all agreeing substantially with each other over their region
of validity.  This suggests that the appropriate units to measure our
results with respect to are 
$T/\sqrt{c_\alpha}$,
which we will do.  Since this rescaling depends on $\alpha$, we can think of $c_\alpha$ as a combination of the massive parameters $c$ and $\phi$.
In figure~\ref{fig:sound2}, we plot our rescaled sound speeds, along with the KTY result (\ref{KTYSound}) and also the function
\begin{eqnarray}
\label{FourthSound}
c_s^2 \approx {1 \over 3 (1 + {c_\alpha^2 \over \pi^2 T^4})} \,,
\end{eqnarray}
which is a better fit to our unified result for the speed of sound.

We can compare our result to the calculation of the speed of sound in other models, such as that for the ${\cal N}=2^*$ theory in \cite{Benincasa:2005iv}.  In terms of conformal symmetry breaking parameters $m_f$ and $m_b$, they found to leading order in each $m_f/T$ and $m_b/T$,
\begin{eqnarray}
\label{TwoStarSound}
c_s = {1 \over \sqrt{3}} \left( 1 - {[\Gamma(3/4)]^2 \over 3 \pi^4} \left( m_f \over T \right)^2 - {1 \over 18 \pi^4} \left(m_b \over T\right)^4 + \ldots \right) \,,
\end{eqnarray}
which broadly matches KTY and our results in the leading behavior: all decrease from the conformal value as the temperature goes down.
For generic values of $m_f$ and $m_b$ (including the ${\cal N}=2$ limit $m_f = m_b$), the $(m_b/T)^4$ term will be subleading, and in this case the ${\cal N}=2^*$ speed of sound matches 
the KTY model
to order $1/T^2$ (including the sign) as long as we identify the symmetry-breaking parameters $c = 2 [\Gamma(3/4)]^2 m_f^2/3 \pi^2$.  
In the limit $m_f \ll m_b$, however (including $m_f = 0$), the ${\cal N}=2^*$ models predict the first subleading term to be proportional to $- 1/T^4$, which is a better match to our models and the form (\ref{FourthSound}).
It is possible that the best fit to our models contains both terms at leading order, as in (\ref{TwoStarSound}).\footnote{Other theories that approach $c_s^2 = 1/3$ from below in the high-$T$ limit 
include the model of \cite{Andreev:2007zv, Andreev:2006eh} which is closely related to KTY,
the finite-temperature Klebanov-Strassler cascade \cite{Aharony:2005zr, Buchel:2005cv}, the Gubser-Nellore model \cite{Gubser1}, and the G\"ursoy-Kiritsis-Mazzanti-Nitti solutions \cite{Gursoy:2008bu, Gursoy:2008za}.   Models with constant $c_s^2 < 1/3$ also exist, including the Sakai-Sugimoto model \cite{Benincasa:2006ei}, the Chamblin-Reall models \cite{Gubser1} and the model of Springer \cite{Springer:2008js, Springer:2009wj}.}

We have seen that as long as we use the right units, our determination of the speed
of sound is quite robust,
and broadly similar to the KTY result though not of precisely the same form.
As a consequence, we can use the physical quantity of the speed of sound to circumvent our ignorance about the physical meanings of $c$ and $\phi$ and use the running of the speed of sound itself as our effective scale.  Let us now examine a worldsheet-dependent quantity, the jet-quenching parameter $\hat{q}$.

\section{Jet-quenching parameter}
\label{sec:jet}

The authors of \cite{LRS} calculate a number of observables associated to the propagation of quarks in the plasma defined by the KTY model, which on the gravity side are associated to worldsheet calculations.  Here we subject our class of models to the calculation of one of these, the jet-quenching parameter $\hat{q}$.

The parameter is computed via the integral\footnote{The definition
  used by \cite{LRS,Buchel:2006bv,Armesto:2006zv},
as proposed by \cite{Liu:2006ug,Liu:2006he},
 has been called into question by \cite{argyres}; since our primary motivation is to compare to the \cite{LRS} computation we will use their definition.}
\begin{equation}\label{eq:qhat}
\hat{q} = \frac{1}{\pi\alpha'}\left(\int_0^{\Phi_h} \frac{d\Phi}{\sqrt{\bar{h}(\Phi)[1-\bar{h}(\Phi)]\exp[6\bar{A}(\Phi)-2\bar{B}(\Phi)]}}\right)^{-1}
\end{equation}
where $\alpha'$ is the string tension, and the bars above the metric
functions are reminders that they must be transformed into string
frame.  We can calculate this quantity for our model as well.  Note that to do this, we must make a choice as to whether $\Phi$ corresponds to the dilaton or not, since (\ref{eq:qhat}) is sensitive to the string metric.  Seeking to have things all ways, we will try both choices and compare.

\begin{figure}
\begin{center}
\includegraphics[width=0.6\textwidth]
{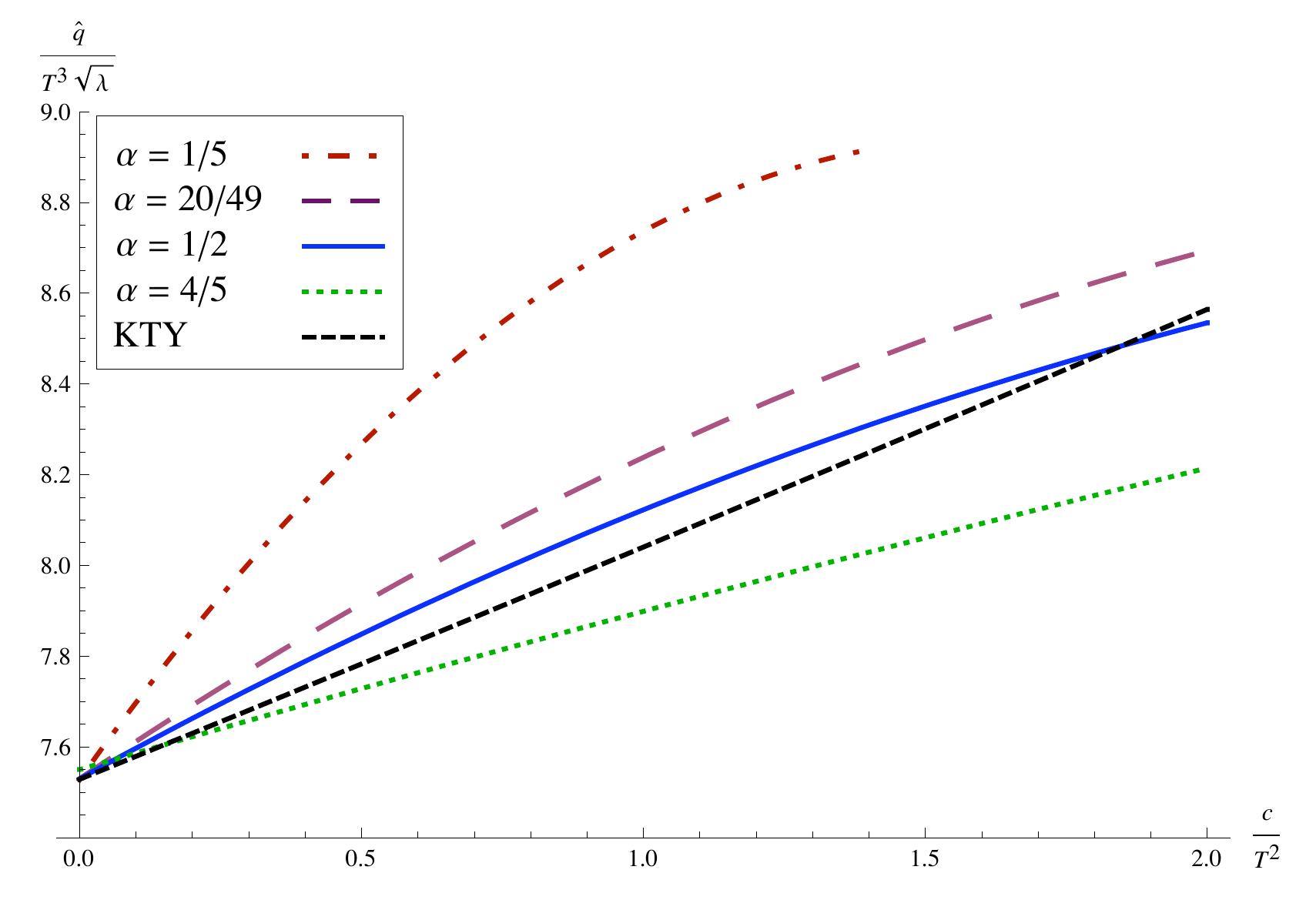}
\caption{Jet-quenching parameter plotted against $c/T^2$ for several
  values of $\alpha$, and in the KTY model. The definition of the
  quantity $\lambda$ follows \cite{LRS}.
\label{fig:qhat-dil-unscaled}}
\end{center}
\end{figure}

First we calculate $\hat{q}$ for various values of $\alpha$ assuming the scalar is the dilaton, and plot it with respect to $T/\sqrt{c}$ in figure~\ref{fig:qhat-dil-unscaled}, along with the value obtained by \cite{LRS}. 
Following \cite{LRS}, we show $\hat{q}$ relative to $T^3 \sqrt{\lambda}$, taking into account that many energy loss models used to describe jet quenching at
RHIC take $\hat{q}$ to scale like $T^3$ \cite{Bass:2008rv}.
 We notice a few things: first of all, the high-temperature (conformal) limit is robust, with the KTY model and our solutions for various $\alpha$ all approaching the same value at $\hat{q}/( T^3 \sqrt{\lambda}) \approx 7.5$.  However, the functional forms of the various values of $\alpha$ diverge substantially as conformal symmetry breaking is introduced,
with small $\alpha$ possessing the greatest slope.

We established in the previous subsection, however, that a more physical scale is to measure quantities relative to $T/\sqrt{c_\alpha}$, since in this scale, every value of $\alpha$ has the same speed of sound.  Thus if the jet-quenching parameter varies when measured relative to $T/\sqrt{c_\alpha}$, this represents a physical difference between models and not just an unknown and arbitrary rescaling of the reference scale.

\begin{figure}
\begin{center}
\includegraphics[width=0.6\textwidth]
{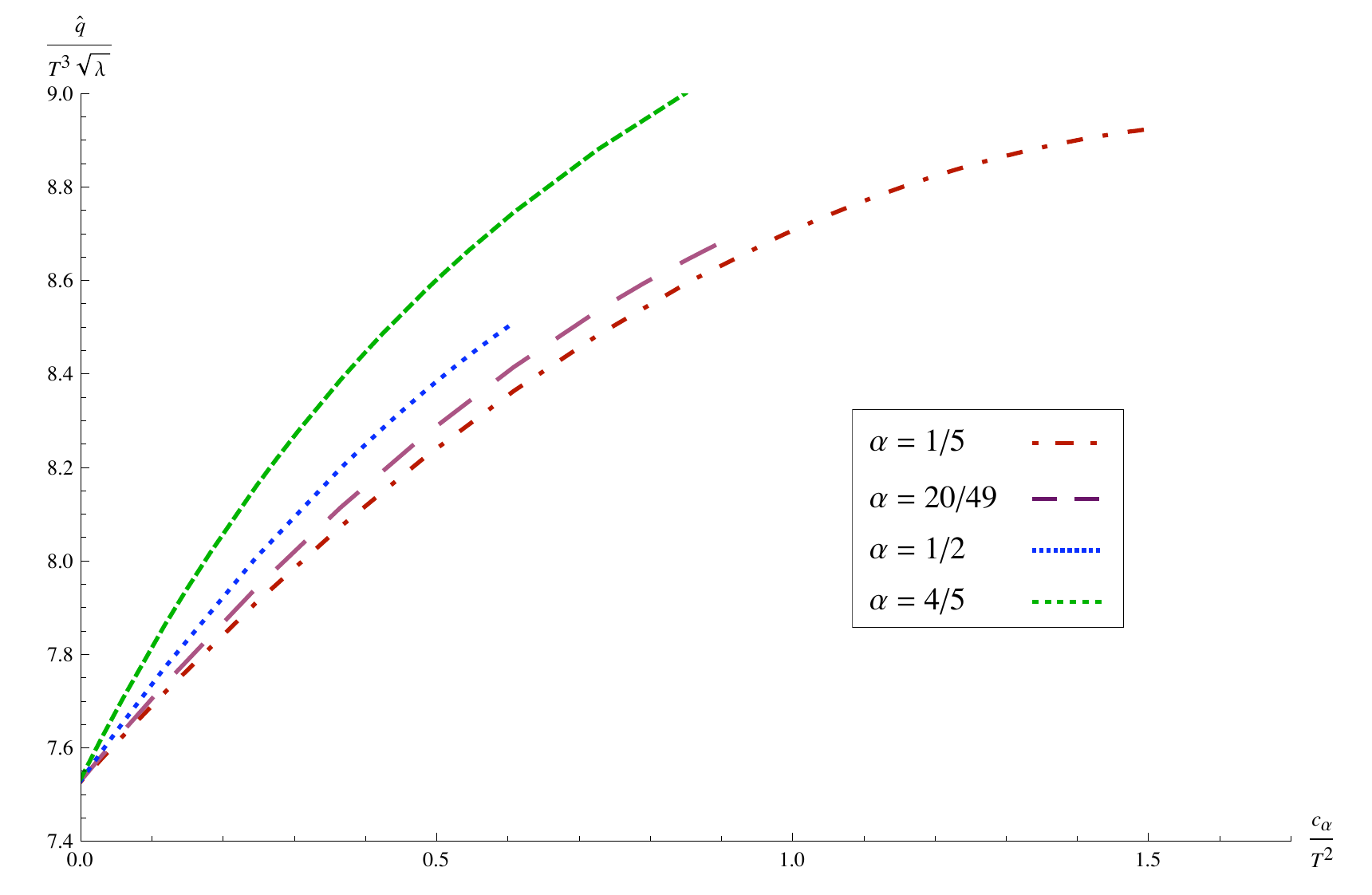}
\caption{ Jet-quenching parameter in scaled coordinates. The 
universality
  observed in the speed of sound calculation is no longer present; our
  model does not seem to uniquely determine the temperature dependence of $\hat{q}$.
\label{fig:qhat-dil-scaled}}
\end{center}
\end{figure}

In figure~\ref{fig:qhat-dil-scaled} we plot the same results relative to the physical scale.  Interestingly, we still find distinct curves for $\hat{q}$ at different values of $\alpha$.  Thus the variation of $\alpha$ represents a true physical deformation, allowing us to vary the functional form of the jet quenching while keeping the ``bulk" property of the speed of sound fixed.  
Notably,
the relative positions of the different values of $\alpha$ have flipped,
with large $\alpha$ now having greater slope.

\begin{figure}
\begin{center}
\includegraphics[width=0.6\textwidth]
{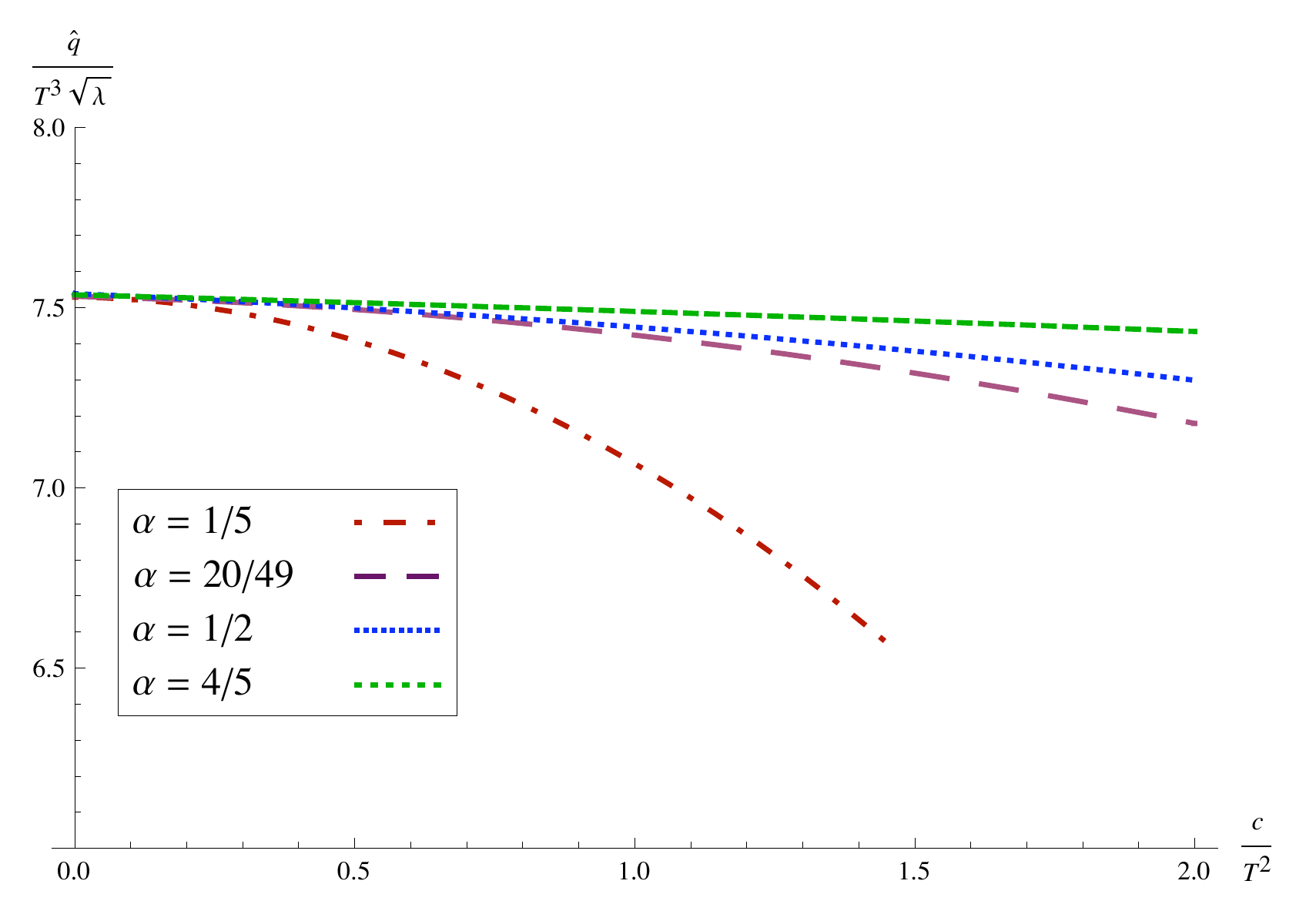}
\caption{Jet-quenching parameter when $\Phi$ is not the dilaton, in unscaled coordinates.
\label{fig:qhat-nodil-unscaled}}
\end{center}
\end{figure}

\begin{figure}
\begin{center}
\includegraphics[width=0.6\textwidth]
{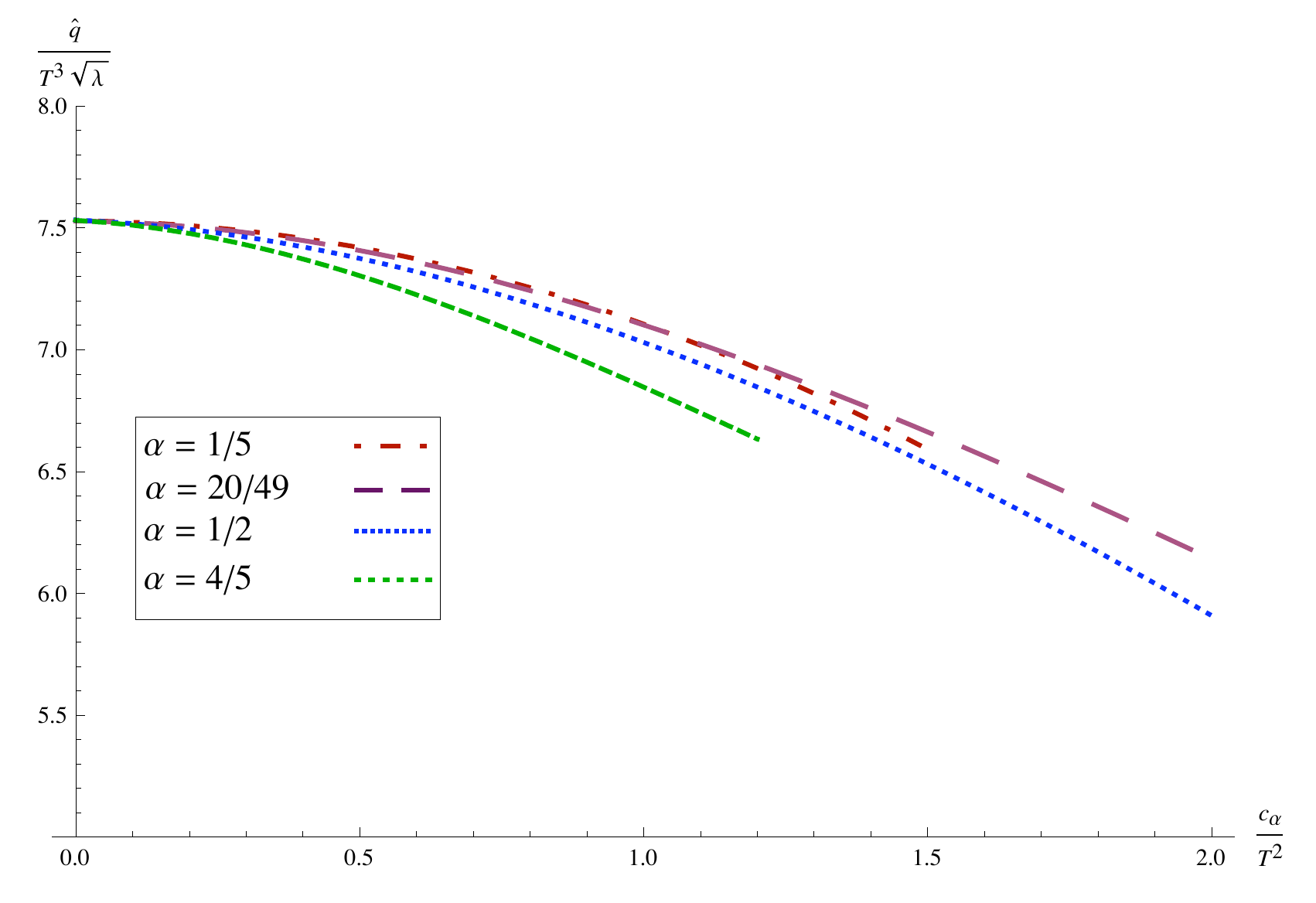}
\caption{Jet-quenching parameter when $\Phi$ is not the dilaton, in scaled coordinates.
\label{fig:qhat-nodil-scaled}}
\end{center}
\end{figure}

We can compare this result to the case where the scalar $\Phi$ is not treated as the dilaton, so the string metric warp factors are identical to the Einstein cases (\ref{eq:A}), (\ref{eq:B}); the results are plotted in figures~\ref{fig:qhat-nodil-unscaled}, \ref{fig:qhat-nodil-scaled}.  We note that the functional form of $\hat{q}$ changes substantially between the two  cases.  For a running dilaton, regardless of $\alpha$ the ratio $\hat{q}/(T^3 \sqrt{\lambda})$ increased as the temperature dropped; for the absence of a running dilaton we find the opposite, with the 
ratio generally falling, more pronouncedly in the scaled (physical) coordinates.  Again the switch to scaled coordinates $T/\sqrt{c_\alpha}$ also exchanges the ordering of the models. 
We discuss these results in the final section.

\section{Discussion and conclusions}
\label{sec:Conclusion}

We set out to try and understand whether the use of simple model geometries not known to solve any equation of motion but possessing a key desired characteristic was reasonable and robust.  At this point we can attempt to address these questions.

There is no indication whatsoever that the use of the KTY model leads
to unphysical results, or indeed results particularly at odds with the
class of solutions to the scalar/gravity system we construct.  For
both the speed of sound and the jet-quenching parameter, the KTY model
makes predictions solidly in the middle of the range of results of our
solutions, and with similar
functional forms.  So
this is reassuring: if there is a hidden difficulty with such a simple
model, we have not found it.

Moreover, we have found that the functional form of the speed of sound is universal across our models, regardless of our tunable parameter $\alpha$, as long as we measure it in the right units; it also agrees with the leading-order behavior of a 
class of ${\cal N} = 2^*$ models \cite{Benincasa:2005iv}.  Thus the predictions of the sound speed are independent of whichever representative of our solutions 
one chooses to use, and hence can be said to be significantly robust.  This does not imply that they are universal across all non-conformal plasmas, 
but are part of a broader class including the KTY model and others that approaches the conformal value from below as the temperature gets large.

For examining the results for the jet-quenching parameter $\hat{q}$, it is useful for us to distinguish two different notions of ``robustness".  As a first point of comparison, we can look at the $\hat{q}$ results
in the context of the results of Liu, Rajagopal and Yee \cite{LRS}. In this work, a dominant thrust was to
identify whether or not a given result was ``robust'' to the
introduction of non-conformality. The idea is that a ``robust''
quantity would change its value little as the conformal symmetry
breaking parameter
relative to the temperature
 was varied across some physically interesting
range. Although the variation of the potential does not allow us to
scan an arbitrary range of, say, $c/T^2$, we can nevertheless 
make an comparison of most of our models over the range considered in \cite{LRS}. 

For the $\alpha$ values 20/49, 1/2 and 4/5, we can
follow $\hat{q}$ all the way from $c/T^2=0$ to 4. Over this range, one finds that
the value of $\hat{q}/ T^3 \sqrt{\lambda}$ changes by about 15\% for
all
$\alpha$ values. By comparison, over the same range the KTY
result 
was found in \cite{LRS} to increase 
by about 28\%. From figure 
\ref{fig:qhat-dil-unscaled} it is easy to see that the largest
variation is expected from $\alpha = 1/5$. Unfortunately the variation
of the potential limits us in this case to the range $0\le c/T^2\le 1.4$, across
which $\hat{q}$ changes by 18\%. When the scalar field is
{\it not} the dilaton, 
figure \ref{fig:qhat-nodil-unscaled}, the jet quenching parameter will
again change most when $\alpha = 1/5$ --- here 13 \% for $0\le c/T^2\le
1.4$. When $\alpha =4/5, 1/2$ or $20/49$, we can explore the range $0\le c/T^2\le
4$, for which we find a 5\%, 10\%, and 10\% change respectively. Accordingly,
over these ranges in $c/T^2$, the value of $\hat{q}$ for a given
$\alpha$ is relatively robust to the introduction of the conformal
symmetry breaking parameter $c$
following the definition of \cite{LRS}.

There is another important way in which our results may be judged for
``robustness''. 
It is evident that none of the geometries we have 
studied is
QCD itself, so their predictions are valuable exactly as far as they are not particularly dependent on the details of the model picked,
as we have discussed. 
The ideal would be to pick a simple model with a few constraints --- no conformal symmetry, the right kind of matter, and so on --- and hope that the physical results (for at least some quantities) would then be independent of 
details of which model we have picked --- in our case, of which value of the dimensionless parameter $\alpha$ was used.
We found that the choice of $\alpha$ could be used to fix a unit system natural for the bulk property of the speed of sound, so  we can normalize the variation of the sound curve across all models.  Even with this normalization, however, the jet-quenching parameter had substantial variation even over the narrow window of $\alpha$-values we were able to explore.

Thus, on the one hand we find that the speed of sound is quite
resilient.
All of our models 
lead to the same functional form for the sound speed over their regions of
validity, which can be made to coincide in the proper set of
units; they closely resemble the model of KTY \cite{Kajantie} as well.

On the other hand, 
although our calculation of the jet-quenching parameter seems quite
physical, it is less robust in this second sense, at least 
away from the conformal limit, as its functional form varies significantly with $\alpha$.
Without knowing the true field-theory dual, it is impossible to know what $\alpha$ corresponds to, and hence which value should be picked,
and therefore which curve to use.
Moreover, 
we could have chosen a more complicated starting point 
for our ansatz
with more dialable parameters, and there is no reason to think the variation of $\hat{q}$ would not also persist over such a multi-dimensional parameter space.

Thus if we want to extract universal properties of something like the
jet-quenching parameter, we either have to provide new input
constraints into our models, or look for a property insensitive to
fine details, caring only about a gross behavior of the quantity.
One example of the latter might be the substantial difference in the
behavior of $\hat{q}$ between when $\Phi$ was treated as the dilaton
and not.  Although the details of the curves vary, one could try to
extract an overall lesson that a running dilaton leads to a growth of
$\hat{q}/ T^3 \sqrt{\lambda}$ as the temperature diminishes, while a
constant dilaton leads to the opposite.  
Currently, many energy loss models used to describe jet quenching at
RHIC take $\hat{q}$ to scale like $T^3$ \cite{Bass:2008rv}, giving no
preference for the fate of our scalar field. Determining higher order
temperature dependence would constitute another input into future
models of the plasma, and place constraints on the available freedoms.

In conclusion, we have found no reason to avoid simple models of the
QCD plasma, but we have reinforced that the robustness of their
predictions must be handled with care.  Some quantities such as the
sound speed may be quite robust, while others such as the
jet-quenching parameter 
may be somewhat robust over the variation of a single parameter but less so once additional dialable parameters are introduced to the family of models.
Even so, they may
contain useful information such as the relationship between the
dilaton and the 
global behavior of the function $\hat{q}(T)$. 
Further study of these questions in the future is surely warranted.

\section*{Acknowledgments}

We are grateful for discussions with Max Brown, Shanta de Alwis, Tom
DeGrand,  Hong Liu, Jamie Nagle and Krishna Rajagopal.  O.D.~thanks
the Aspen Center for Physics where this work was initiated.  
C.R.~thanks the organizers of the Hot Quarks 2008 conference where
portions of this work were originally presented.
This work was supported by the DOE under grant DE-FG02-91-ER-40672.


\begin{thebibliography}{99}

\bibitem{son}
  P.~Kovtun, D.~T.~Son and A.~O.~Starinets,
  ``Viscosity in strongly interacting quantum field theories from black hole
  physics,''
  Phys.\ Rev.\ Lett.\  {\bf 94}, 111601 (2005)
  [arXiv:hep-th/0405231].



\bibitem{Kats:2007mq}
  Y.~Kats and P.~Petrov,
  ``Effect of curvature squared corrections in AdS on the viscosity of the dual
  gauge theory,''
  JHEP {\bf 0901}, 044 (2009)
  [arXiv:0712.0743 [hep-th]].

\bibitem{Brigante:2007nu}
  M.~Brigante, H.~Liu, R.~C.~Myers, S.~Shenker and S.~Yaida,
  ``Viscosity Bound Violation in Higher Derivative Gravity,''
  Phys.\ Rev.\  D {\bf 77}, 126006 (2008)
  [arXiv:0712.0805 [hep-th]].

\bibitem{adams}
  A.~Adams, A.~Maloney, A.~Sinha and S.~E.~Vazquez,
  ``1/N Effects in Non-Relativistic Gauge-Gravity Duality,''
  arXiv:0812.0166 [hep-th].

\bibitem{Buchel:2008vz}
  A.~Buchel, R.~C.~Myers and A.~Sinha,
  ``Beyond eta/s = 1/4pi,''
  arXiv:0812.2521 [hep-th].


\bibitem{Gubser:2009md}
  S.~S.~Gubser and A.~Karch,
  ``From gauge-string duality to strong interactions: a Pedestrian's Guide,''
  arXiv:0901.0935 [hep-th].

\bibitem{Erdmenger:2007cm}
  J.~Erdmenger, N.~Evans, I.~Kirsch and E.~Threlfall,
  ``Mesons in Gauge/Gravity Duals - A Review,''
  Eur.\ Phys.\ J.\  A {\bf 35}, 81 (2008)
  [arXiv:0711.4467 [hep-th]].

\bibitem{Mateos:2007ay}
  D.~Mateos,
  ``String Theory and Quantum Chromodynamics,''
  Class.\ Quant.\ Grav.\  {\bf 24}, S713 (2007)
  [arXiv:0709.1523 [hep-th]].


\bibitem{buchel}
  A.~Buchel,
  ``N = 2* hydrodynamics,''
  Nucl.\ Phys.\  B {\bf 708}, 451 (2005)
  [arXiv:hep-th/0406200].

\bibitem{Buchel:2008uu}
  A.~Buchel and C.~Pagnutti,
  ``Bulk viscosity of N=2* plasma,''
  arXiv:0812.3623 [hep-th].

\bibitem{Buchel:2007vy}
  A.~Buchel, S.~Deakin, P.~Kerner and J.~T.~Liu,
  ``Thermodynamics of the N = 2* strongly coupled plasma,''
  Nucl.\ Phys.\  B {\bf 784}, 72 (2007)
  [arXiv:hep-th/0701142].


\bibitem{sak-sug}
  T.~Sakai and S.~Sugimoto,
  ``Low energy hadron physics in holographic QCD,''
  Prog.\ Theor.\ Phys.\  {\bf 113}, 843 (2005)
  [arXiv:hep-th/0412141].

\bibitem{Sakai:2005yt}
  T.~Sakai and S.~Sugimoto,
  ``More on a holographic dual of QCD,''
  Prog.\ Theor.\ Phys.\  {\bf 114}, 1083 (2005)
  [arXiv:hep-th/0507073].

\bibitem{Aharony:2006da}
  O.~Aharony, J.~Sonnenschein and S.~Yankielowicz,
  ``A holographic model of deconfinement and chiral symmetry restoration,''
  Annals Phys.\  {\bf 322}, 1420 (2007)
  [arXiv:hep-th/0604161].

\bibitem{Parnachev:2006dn}
  A.~Parnachev and D.~A.~Sahakyan,
  ``Chiral phase transition from string theory,''
  Phys.\ Rev.\ Lett.\  {\bf 97}, 111601 (2006)
  [arXiv:hep-th/0604173].
  
\bibitem{Kajantie}
  K.~Kajantie, T.~Tahkokallio and J.~T.~Yee,
  ``Thermodynamics of AdS/QCD,''
  JHEP {\bf 0701}, 019 (2007)
  [arXiv:hep-ph/0609254].


\bibitem{LRS}
  H.~Liu, K.~Rajagopal and Y.~Shi,
  ``Robustness and Infrared Sensitivity of Various Observables in the
  Application of AdS/CFT to Heavy Ion Collisions,''
  JHEP {\bf 0808}, 048 (2008)
  [arXiv:0803.3214 [hep-ph]].


\bibitem{Gubser1}
  S.~S.~Gubser and A.~Nellore,
  ``Mimicking the QCD equation of state with a dual black hole,''
  Phys.\ Rev.\  D {\bf 78}, 086007 (2008)
  [arXiv:0804.0434 [hep-th]].

\bibitem{Gubser2}
 S.~S.~Gubser, S.~S.~Pufu and F.~D.~Rocha,
  ``Bulk viscosity of strongly coupled plasmas with holographic duals,''
  JHEP {\bf 0808}, 085 (2008)
  [arXiv:0806.0407 [hep-th]].

\bibitem{Gubser:2008yx}
  S.~S.~Gubser, A.~Nellore, S.~S.~Pufu and F.~D.~Rocha,
  ``Thermodynamics and bulk viscosity of approximate black hole duals to finite
  temperature quantum chromodynamics,''
  Phys.\ Rev.\ Lett.\  {\bf 101}, 131601 (2008)
  [arXiv:0804.1950 [hep-th]].


\bibitem{Gursoy:2008bu}
  U.~Gursoy, E.~Kiritsis, L.~Mazzanti and F.~Nitti,
  ``Deconfinement and Gluon Plasma Dynamics in Improved Holographic QCD,''
  Phys.\ Rev.\ Lett.\  {\bf 101}, 181601 (2008)
  [arXiv:0804.0899 [hep-th]].

\bibitem{Gursoy:2008za}
  U.~Gursoy, E.~Kiritsis, L.~Mazzanti and F.~Nitti,
  ``Holography and Thermodynamics of 5D Dilaton-gravity,''
  arXiv:0812.0792 [hep-th].


\bibitem{Andreev:2007zv}
  O.~Andreev,
  ``Some Thermodynamic Aspects of Pure Glue, Fuzzy Bags and Gauge/String
  Duality,''
  Phys.\ Rev.\  D {\bf 76}, 087702 (2007)
  [arXiv:0706.3120 [hep-ph]].
  
  \bibitem{Andreev:2006eh}
  O.~Andreev and V.~I.~Zakharov,
  ``The Spatial String Tension, Thermal Phase Transition, and AdS/QCD,''
  Phys.\ Lett.\  B {\bf 645}, 437 (2007)
  [arXiv:hep-ph/0607026].

\bibitem{Muller:2006ee}
  B.~Muller and J.~L.~Nagle,
  ``Results from the Relativistic Heavy Ion Collider,''
  Ann.\ Rev.\ Nucl.\ Part.\ Sci.\  {\bf 56}, 93 (2006)
  [arXiv:nucl-th/0602029].

\bibitem{Benincasa:2005iv}
  P.~Benincasa, A.~Buchel and A.~O.~Starinets,
  ``Sound waves in strongly coupled non-conformal gauge theory plasma,''
  Nucl.\ Phys.\  B {\bf 733}, 160 (2006)
  [arXiv:hep-th/0507026].


\bibitem{Aharony:2005zr}
  O.~Aharony, A.~Buchel and A.~Yarom,
  ``Holographic renormalization of cascading gauge theories,''
  Phys.\ Rev.\  D {\bf 72}, 066003 (2005)
  [arXiv:hep-th/0506002].


\bibitem{Buchel:2005cv}
  A.~Buchel,
  ``Transport properties of cascading gauge theories,''
  Phys.\ Rev.\  D {\bf 72}, 106002 (2005)
  [arXiv:hep-th/0509083].

\bibitem{Benincasa:2006ei}
  P.~Benincasa and A.~Buchel,
  ``Hydrodynamics of Sakai-Sugimoto model in the quenched approximation,''
  Phys.\ Lett.\  B {\bf 640}, 108 (2006)
  [arXiv:hep-th/0605076].



\bibitem{Springer:2008js}
  T.~Springer,
  ``Sound Mode Hydrodynamics from Bulk Scalar Fields,''
  Phys.\ Rev.\  D {\bf 79}, 046003 (2009)
  [arXiv:0810.4354 [hep-th]].

\bibitem{Springer:2009wj}
  T.~Springer,
  ``Second order hydrodynamics for a special class of gravity duals,''
  arXiv:0902.2566 [hep-th].



\bibitem{Liu:2006ug}
  H.~Liu, K.~Rajagopal and U.~A.~Wiedemann,
  ``Calculating the jet quenching parameter from AdS/CFT,''
  Phys.\ Rev.\ Lett.\  {\bf 97}, 182301 (2006)
  [arXiv:hep-ph/0605178].

\bibitem{Liu:2006he}
  H.~Liu, K.~Rajagopal and U.~A.~Wiedemann,
  ``Wilson loops in heavy ion collisions and their calculation in AdS/CFT,''
  JHEP {\bf 0703}, 066 (2007)
  [arXiv:hep-ph/0612168].

\bibitem{Buchel:2006bv}
  A.~Buchel,
  ``On jet quenching parameters in strongly coupled non-conformal gauge
  theories,''
  Phys.\ Rev.\  D {\bf 74}, 046006 (2006)
  [arXiv:hep-th/0605178].

\bibitem{Armesto:2006zv}
  N.~Armesto, J.~D.~Edelstein and J.~Mas,
  ``Jet quenching at finite 't Hooft coupling and chemical potential from
  AdS/CFT,''
  JHEP {\bf 0609}, 039 (2006)
  [arXiv:hep-ph/0606245].

\bibitem{argyres}
  P.~C.~Argyres, M.~Edalati and J.~F.~Vazquez-Poritz,
  ``Lightlike Wilson loops from AdS/CFT,''
  JHEP {\bf 0803}, 071 (2008)
  [arXiv:0801.4594 [hep-th]].


\bibitem{Bass:2008rv}
  S.~A.~Bass, C.~Gale, A.~Majumder, C.~Nonaka, G.~Y.~Qin, T.~Renk and J.~Ruppert,
  ``Systematic Comparison of Jet Energy-Loss Schemes in a realistic
  hydrodynamic medium,''
  arXiv:0808.0908 [nucl-th].




\end{thebibliography}
\end{document}